\documentclass [12pt] {article}
\usepackage[dvips]{graphicx}

\begin{document}

\title{Statistical Approach of Modulational Instability in the Class of 
Derivative Nonlinear Schr\"odinger Equations}
\author{ A.T. Grecu, 
D. Grecu, Anca Visinescu \\
\small{\it Department of Theoretical Physics} \\
\small{\it ``Horia Hulubei'' National Institute of Physics and Nuclear 
Engineering} \\
\small{Bucharest, P.O. Box MG-6, Romania}\\
\small{e-mail: avisin@theor1.theory.nipne.ro}
}
\date{}
\maketitle

\abstract{The modulational instability (MI) in the class of NLS equations 
is discussed using a statistical approach (SAMI). A kinetic equation for 
the two-point correlation function is studied in a linear approximation, 
and an integral stability equation is found. The modulational instability 
is associated with a positive imaginary part of the frequency. The integral 
equation is solved for different types of initial distributions ($\delta$-
function, Lorentzian) and the results are compared with those obtained using 
a deterministic approach (DAMI). The differences between MI of 
the normal NLS equation and derivative NLS equations is emphasized.}



\section{Introduction}

\indent The modulational instability (MI - also known as Benjamin-Feir 
instability) is a well known phenomenon encountered, in quite general 
conditions, almost each time a quasi-monochromatic wave propagates in a 
weakly nonlinear medium (Dodd et al 1982, Remoissenet 1999, Scott 
2003, Benjamin and Feir 1967, Bespalov et al 1966 and Abdullaev et 
al 2002).
It is responsible for the generation of robust, localized excitations, 
the solitons in completely integrable 
systems, and solitary waves in non-integrable ones.

Two different ways to discuss this phenomenon are possible. The first 
one is a deterministic approach (DAMI) in which the amplitude of a plane 
wave, with an amplitude dependent dispersion relation (a Stokes wave), 
is slowly modulated 

\begin{equation} \label{eq1}
\Psi(x,t)=a\left(1+\varepsilon A(x,t)\right)\exp\left(\mathbf{i}
(kx-\omega t)\right)
\end{equation}
\noindent This approach is well known and can be found in any textbook 
on nonlinear physics (see Dodd et al 1982, Remoissenet 1999). 
The second approach is a statistical one (SAMI), and its aim is to see 
the influence of the statistical properties of the medium on the MI 
phenomenon. It is a less used approach, but in the last ten years was 
intensively applied to investigate different physical situations 
(Alber 1978, Janssen 1983, Fedele et al. 2000, Hall et al 2002, 
Lisak et al 2002, Fedele et al 2002, Onorato et al 2003), ranging 
from hydrodynamics (Janssen 1983, Fedele et al. 2000, Hall et al 2002, 
Lisak et al 2002, Fedele et al 2002), to plasma 
physics (Fedele 2002), large amplitude wave-envelope propagation in nonlinear 
media, high-energy accelerators, condensed matter. 
Recently it was extended also to discrete systems (Visinescu et al 
2003, Grecu D. et al 2005) and to 
systems of coupled nonlinear equations (Manakov's system) (Grecu D. et al 
2004 and 2005).

In this approach a kinetic equation for the two-point correlation function
\begin{equation} \label{eq2}
\rho(x_1,x_2)=\left<\Psi(x_1)\,\Psi^*(x_2)\right>
\end{equation}
\noindent is written down and a linear stability analysis near an initial 
distribution
\begin{equation} \label{eq3}
\rho_0(x_1,x_2)=\rho_0(\left|x_1-x_2\right|)
\end{equation}
\noindent is performed. The initial state is assumed homogenous and 
consequently $\rho_0$ depends only on 
$|x_1-x_2|$. In the previous expression of $\rho(x_1,x_2)$ by 
$\left<\ldots\right>$ we understand a 
statistical average over the distribution function of the medium.

The aim of this paper is to investigate MI, especially using the statistical 
approach of the following equation
\begin{equation}
\label{eq4}
\mathbf{i} \frac{\partial \Psi}{\partial t} + {1\over 2}\frac{\partial^2 
\Psi}{\partial x^2}+\lambda |\Psi|^2 \Psi +\mathbf{i} \mu |\Psi|^2
\frac{\partial \Psi}{\partial x} +\mathbf{i} \nu \frac{\partial |\Psi|^2}
{\partial x}\Psi=0
\end{equation}
\noindent It is an extended NLS equation, and for different choises of 
the real parameters $\lambda,\mu,\nu$ several completely integrable equations 
are recovered. If $\mu=\nu=0$, we get the well known NLS equation
\begin{equation} \label{eq5}
\mathbf{i} \frac{\partial \Psi}{\partial t} + {1\over 2}\frac{\partial^2 
\Psi}{\partial x^2} + \lambda|\Psi|^2 \Psi=0
\end{equation}
while for $\lambda=0,\mu=\nu$ one obtains the completely integrable 
derivative NLS-I equation (dNLS I) (Kaup 1978, Nakamura et al. 1980, 
Mio et al 1976, Mj\o lhus 1976)
\begin{equation} \label{eq6}
\mathbf{i} \frac{\partial \Psi}{\partial t} + {1\over 2}\frac{\partial^2 
\Psi}{\partial x^2} + \mathbf{i}\mu\frac{\partial}{\partial x}
\left(|\Psi|^2\Psi\right)=0
\end{equation}
and for $\lambda=\nu=0$ we get another integrable derivative NLS II 
equation (dNLS II) (Chen, Lee 1979)
\begin{equation} \label{eq7}
\mathbf{i}\frac{\partial \Psi}{\partial t} + {1\over 2}\frac{\partial^2 
\Psi}{\partial x^2}+\mathbf{i}\mu|\Psi|^2\frac{\partial \Psi}{\partial x}=0
\end{equation}

\begin{figure}[ht]
\label{fig1}
\hfill
\begin{minipage}{.45\textwidth}
\begin{center}  
\includegraphics[scale=0.3]{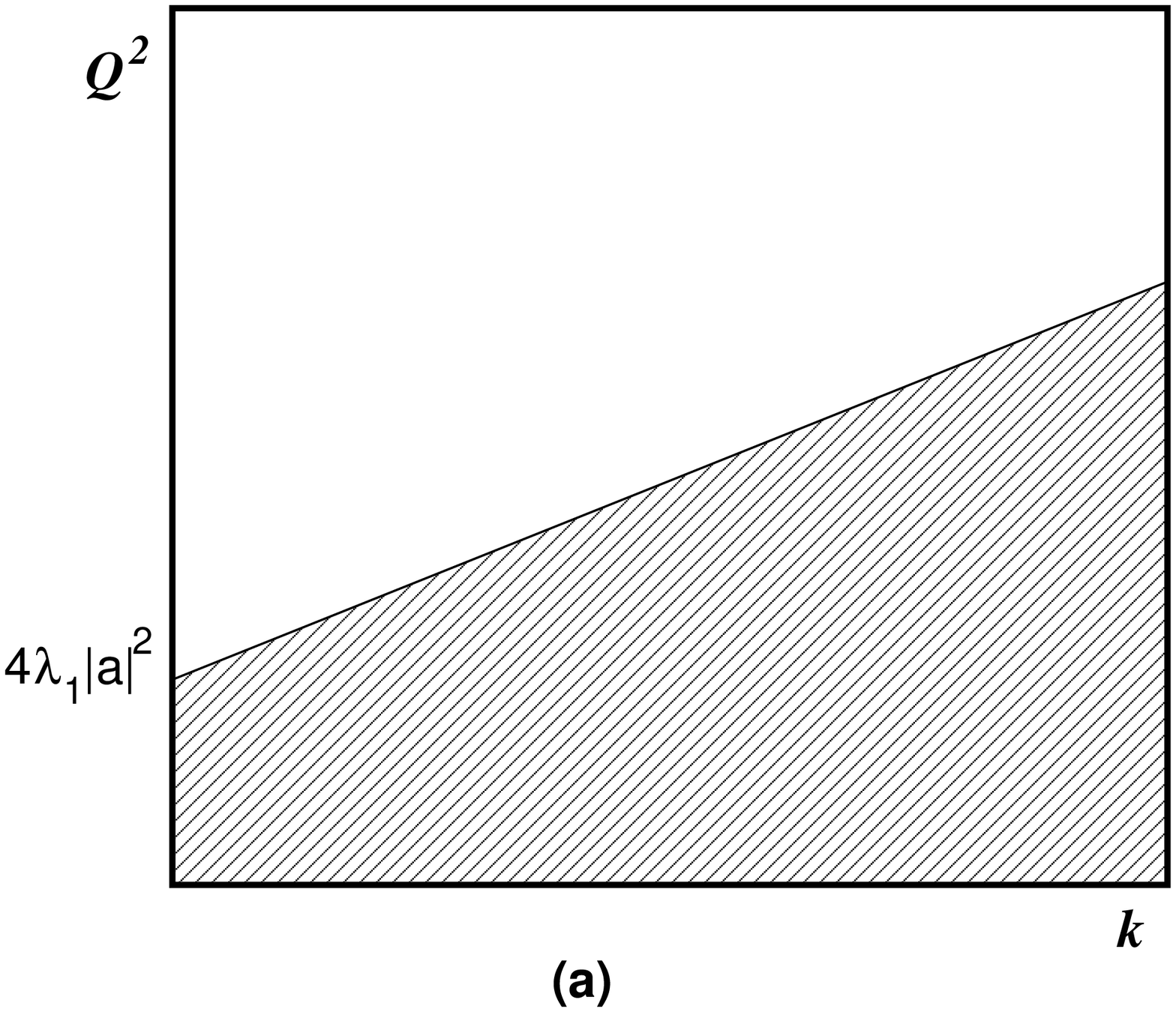}
\end{center}
\end{minipage}
\hfill
\begin{minipage}{.45\textwidth}
\begin{center} 
\includegraphics[scale=0.3]{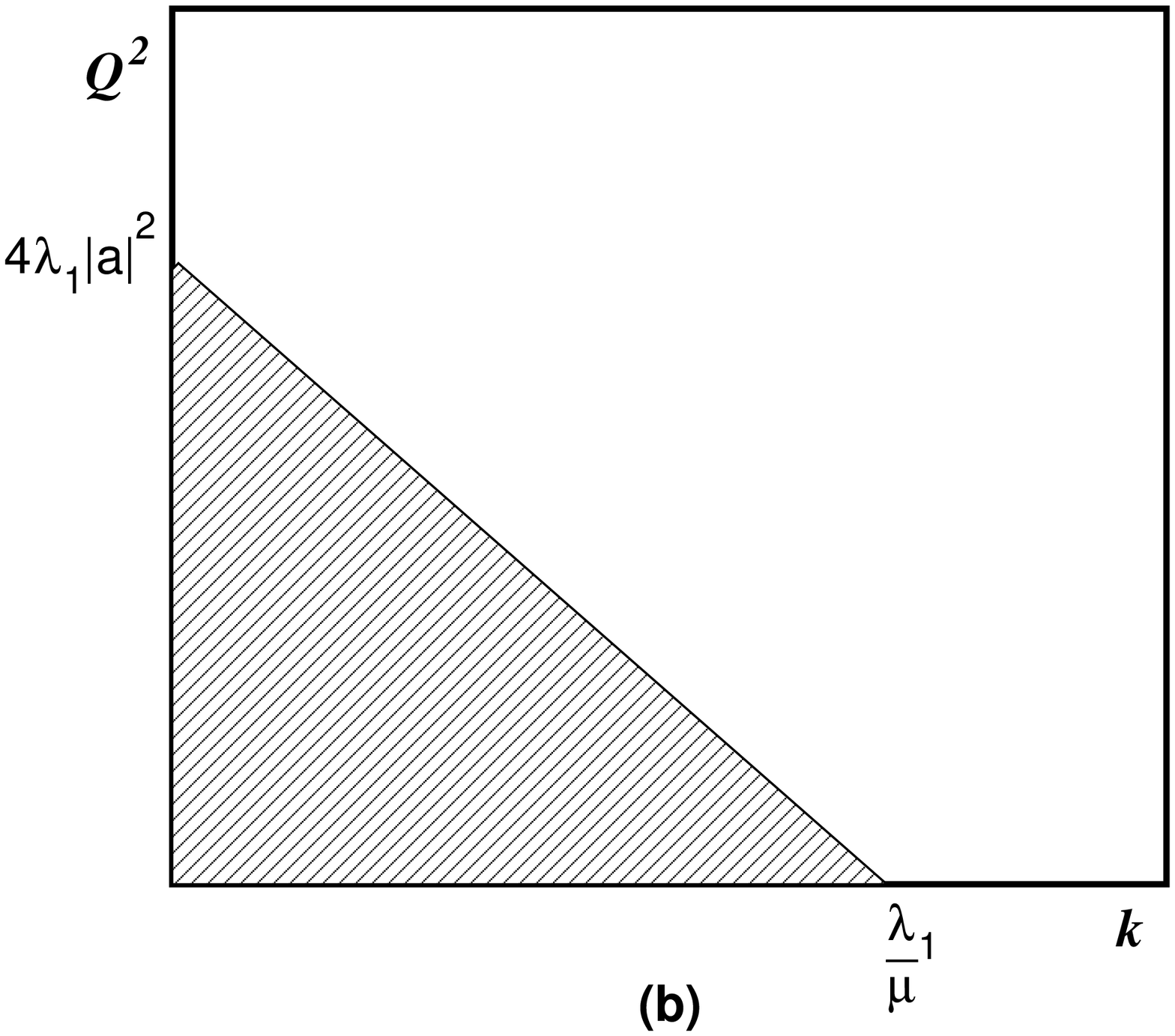}
\end{center}
\end{minipage}
\hfill
\caption{\small Instability regions $(\lambda \neq 0)$; (a) $\mu < 0$, 
(b) $\mu>0$.}
\end{figure}

Before starting the SAMI study of eq. (\ref{eq4}), let us briefly recall 
the results of a DAMI analysis. Introducing (\ref{eq1}) in (\ref{eq4}) and 
keeping only the linear terms in $\varepsilon$ a linear evolution equation for 
$A(x,t)$ is found. Looking for plane wave solutions of it
$$
A(x,t)=\alpha_1\mathrm{e}^{\mathbf{i}(Qx-\Omega t)}+\alpha_{2}^{*}
\mathrm{e}^{-\mathbf{i}(Qx-\Omega^* t)}
$$
the compatibility condition of the homogenous system leads us to
\begin{equation}
\label{eq8}
\Omega-Q\left[k+(\mu+\nu)|a|^2\right]=\mathbf{i} Q\sqrt{(\lambda-
\nu|a|^2-\mu k)|a|^2-\frac{Q^2}{4}}
\end{equation}
an the instability is related to $\mathrm{Im}\Omega>0$. For $\lambda\neq 0$ 
and denoting 
$\lambda_1 = \lambda - \nu |a|^2$ we obtain
\begin{equation} \label{eq9}
\mathrm{Im}\Omega = Q\sqrt{(\lambda_1-\mu k)|a|^2-\frac{Q^2}{4}}
\end{equation}
and one sees that the instability regions depend on the sign of $\mu$. 
In a $(Q^2,\,k)$ representation 
these are given in Fig. \ref{fig1} (arbitrary units)

It is interesting to consider the instability regions for the two 
integrable dNLS equations. For dNLS I an instability arises only if 
$\mu<0$. We have
\begin{equation} \label{eq10}
\mathrm{Im} \Omega = Q\sqrt{|\mu||a|^2(k-|\mu||a|^2)-\frac{Q^2}{4}}
\end{equation}
and in a $(Q^2,\,k)$ representation the instability occurs only if 
$k>|\mu||a|^2$. For dNLS II the same condition, $\mu<0$, must be 
fulfilled and
\begin{equation} \label{eq11}
\mathrm{Im} \Omega=Q\sqrt{|\mu||a|^2 k-\frac{Q^2}{4}}
\end{equation}
The instability regions in $(Q^2,\,k)$ representation (arbitrary units) 
are presented in Fig. 2.
\begin{figure}[ht]
\label{fig2}
\hfill
\begin{minipage}[t]{.45\textwidth}
\begin{center}  
\includegraphics[scale=0.3]{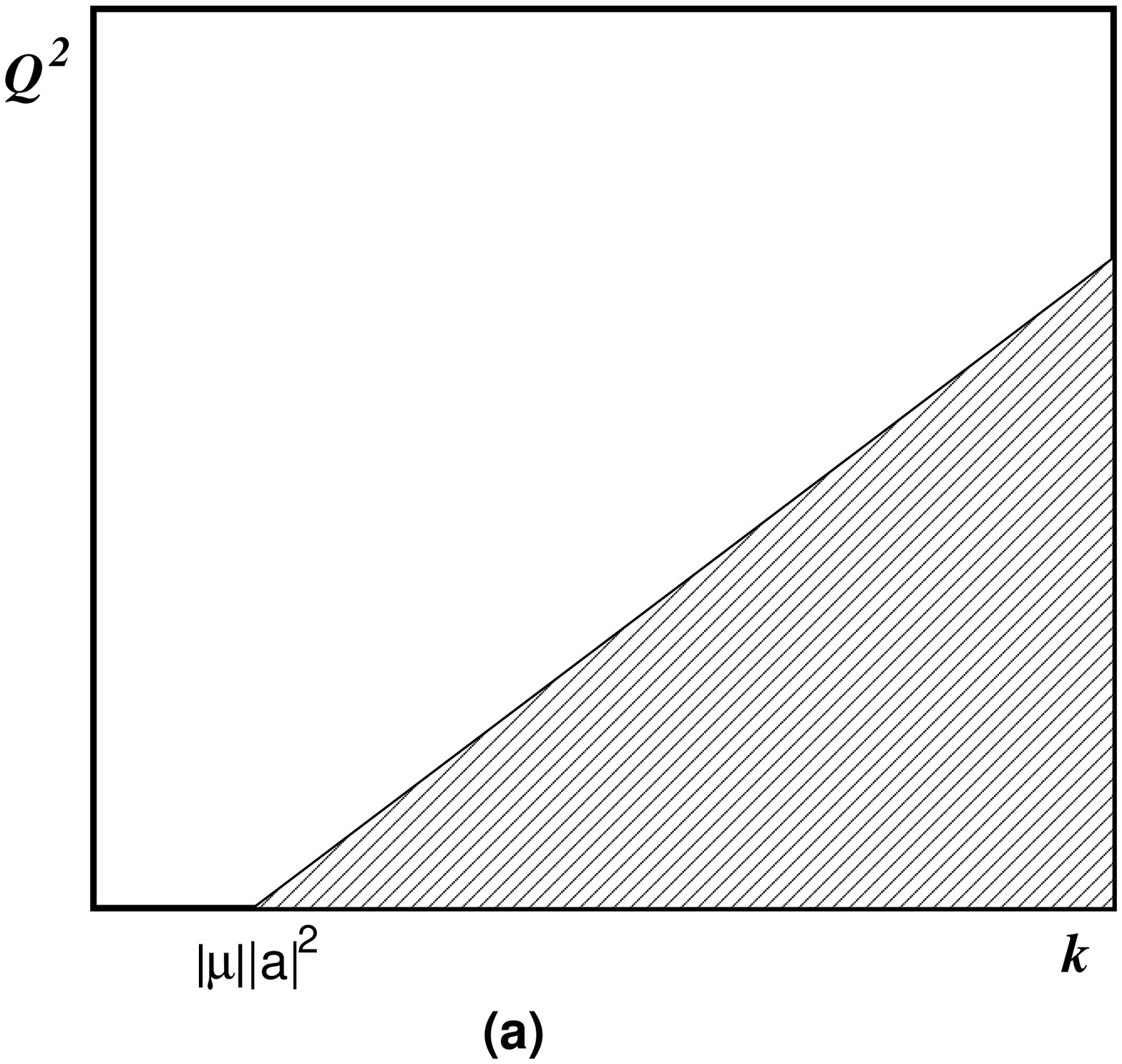}
\end{center}
\end{minipage}
\hfill
\begin{minipage}[t]{.45\textwidth}
\begin{center} 
\includegraphics[scale=0.3]{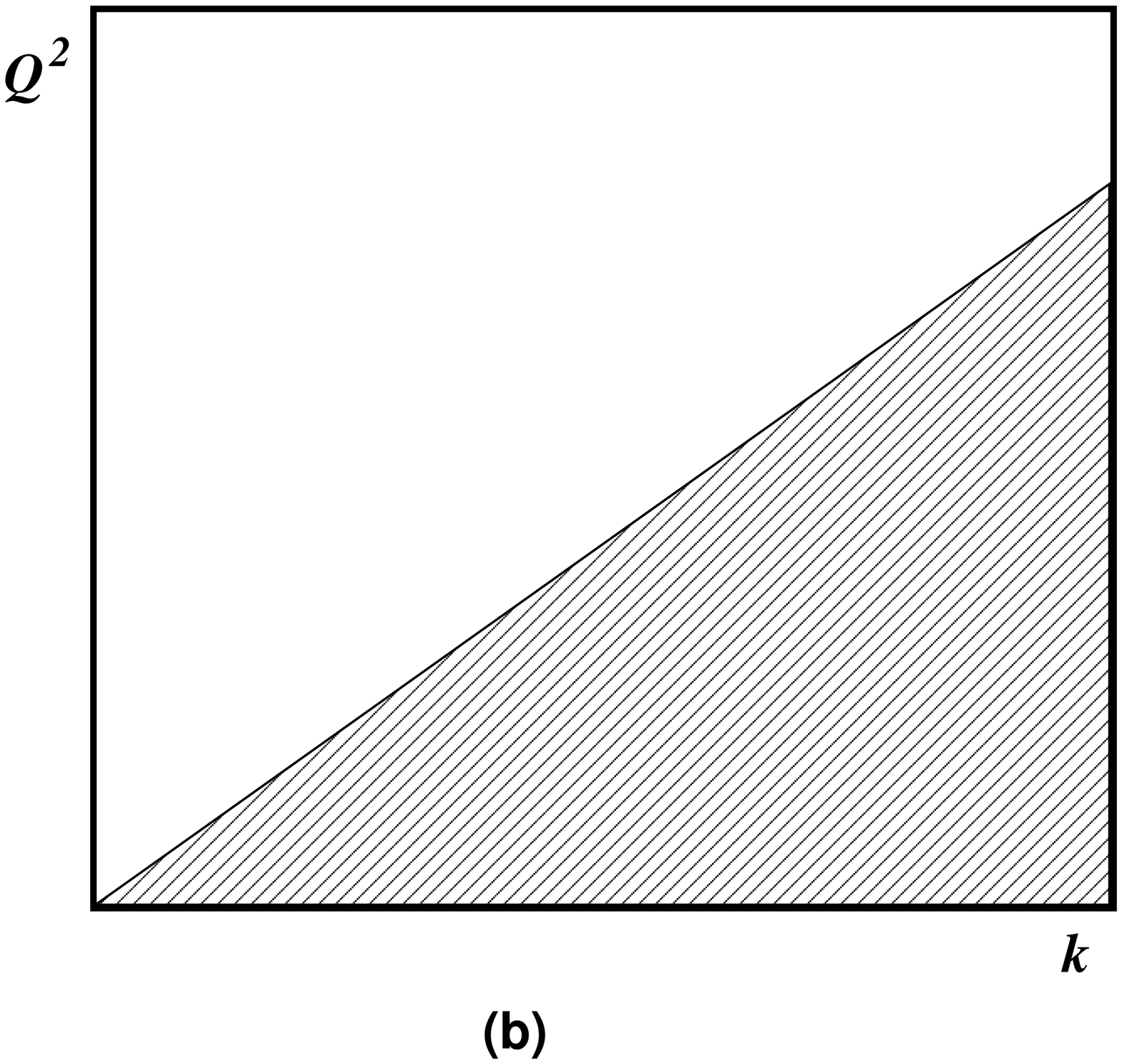}
\end{center}
\end{minipage}
\hfill
\caption{\small Instability regions of the integrable dNLS equations: 
(a) dNLS I, (b) dNLS II}
\end{figure}

\noindent These results have to be compared with the similar result 
obtained for NLS equation (\ref{eq5}), which writes
\begin{equation} \label{eq12}
\mathrm{Im} \Omega=Q\sqrt{\lambda |a|^2 - \frac{Q^2}{4}}
\end{equation}
We see that in the NLS case the instability region is independent on 
the wave vector $k$ of the carrying wave, and restricted to the long 
wave-length region ($Q^2 < 4\lambda |a|^2$), while for all the situations 
discussed above they are $k$-dependent, and more over they are not 
restricted to the long wavelength region (exception is the case (b) 
of Fig 1). Actually for the dNLS equations the long wavelength region 
is either forbidden, or highly unfavourable for the development of the MI.

In the next section the kinetic equation for the two-point correlation 
function (\ref{eq2}) is obtained and its stability is studied in the 
linear approximation. An integral stability equation will be obtained 
and solved for different initial distributions in section 3 
($\delta$-distribution, Lorentzian). Few conclusions and comments are 
given in the last section. Preliminary results have been presented in 
Grecu, A.T. 2005.

\section{Kinetic equation. Linear stability analysis}

The kinetic equation for the two-point correlation function (\ref{eq2}), 
is obtained easily in the following way:
\begin{enumerate}
\item write the equation (\ref{eq4}) for $x=x_1$ and multiply by 
$\Psi^{*}(x_2)$;
\item write the complex conjugated of (\ref{eq4}) for $x=x_2$ and 
multiply by $\Psi(x_1)$;
\item add the two equations and take the statistical average;
\item finally use a Gaussian approximation to decouple the four-point 
correlation functions.
\end{enumerate}
As an illustration we give below two examples of such decouplings
$$\left < \Psi(x_1) \Psi^{*}(x_1) \Psi(x_1) \Psi^*(x_2) \right > = 
2 \overline{a^{2}}(x_1)\rho(x_1,x_2)$$
$$\left < \Psi(x_1) \Psi^*(x_1) \frac{\partial \Psi(x_1)}{\partial x_1} 
\Psi^*(x_2)\right > =$$
$$ =\overline{a^{2}}(x_1) \frac{\partial}{\partial  x_1}\rho(x_1,x_2) +
\rho(x_1,x_2) \lim_{x_2\to x_1} \frac{\partial}{\partial x_1}\rho(x_1,x_2)$$
where
\begin{equation} \label{eq13}
\overline{a^{2}}(x_1)=\left < \Psi(x_1) \Psi^*(x_1) \right >
\end{equation}
The kinetic equation found in this way is
\begin{eqnarray} \label{eq14}
& & \mathbf{i} \frac{\partial \rho}{\partial t} +{1 \over 2}
\left(\frac{\partial^2}{\partial x_{1}^{2}} - \frac{\partial}
{\partial x_{2}^{2}}\right)\rho  + 2 \lambda \left[ \overline{a^{2}}
(x_1)-\overline{a^{2}}(x_2)\right]\rho  \nonumber \\
& & + \mathbf{i} (\mu+\nu)\left[ \overline{a^{2}}(x_1)\frac{\partial}
{\partial x_1} + \overline{a^{2}}(x_2)\frac{\partial}{\partial x_2}\right]
\rho  \\
& & + \mathbf{i} (\mu+\nu)\rho\left[ \lim_{x_2 \to x_1}\frac{\partial}
{\partial x_1} + \lim_{x_1 \to x_2}\frac{\partial}{\partial x_2}\right]\rho
  \nonumber \\
& & + 2\mathbf{i} \nu \rho\left[ \lim_{x_2 \to x_1}\frac{\partial}
{\partial x_1} +\lim_{x_1 \to x_2}\frac{\partial}{\partial x_2}\right]\rho^* = 0 \nonumber
\end{eqnarray}

The next step is to introduce the new set of variables
\begin{eqnarray} \label{eq15}
& & x=x_1-x_2 \;\;\; \mathrm{relative\: coordinate} \\
& & X={1\over 2}(x_1+x_2) \;\; \mathrm{center\: of\: mass\: coordinate}
\nonumber
\end{eqnarray}
and make a Fourier transform with respect to the relative coordinate. 
This procedure is known as the Wigner-Moyal transform (Wigner 1932, 
Moyal 1949) (also 
known as Klimontovich's statistical average method, Toda et al 1995), 
and is very useful to deal with non-homogenous evolution equations. The 
Fourier transform is defined as 
\begin{equation} \label{eq16}
F(k,X,t) = {1\over{2\pi}} \int_{-\infty}^{+\infty}{ \,\mathrm{d}x\, 
\mathrm{e}^{-\mathbf{i} kx} \rho(x,X,t)}
\end{equation}
Using the definition (\ref{eq13}) of $\overline{a^{2}}$, it is easily 
seen that
\begin{equation} \label{eq17}
\overline{a^{2}}(X,t)=\int_{-\infty}^{+\infty}{\,\mathrm{d}k\, F(k,X,t)}
\end{equation}
After straightforward calculations the Fourier transform of (\ref{eq14}) 
is given by
\begin{eqnarray} \label{eq18}
& & \frac{\partial F}{\partial t} + k \frac{\partial F}{\partial X} + 
4\lambda \sum_{j=0}^{\infty}\frac{(-1)^j}{(2j+1)\,!\,2^{2j+1}} 
\frac{\partial^{2j+1} \overline{a^{2}}(X)}{\partial X^{2j+1}}
\frac{\partial^{2j+1} F(k,X)}{\partial k^{2j+1}} \nonumber \\
& & + (\mu+\nu) \frac{\partial \overline{a^{2}}(X)}{\partial X}F(k,X)+ 
2\nu \frac{\partial \left[\overline{a^{2}}(X)\right]^*}{\partial X}F(k,X) \\
& & + (\mu+\nu) \sum_{j=0}^{\infty}{\frac{(-1)^j}{2j\,! 2^{2j}}
\frac{\partial^{2j} \overline{a^{2}}(X)}{\partial X^{2j}}\frac{\partial}
{\partial X}\frac{\partial^{2j} F(k,X)}{\partial k^{2j}}} \nonumber \\
& & - 2(\mu+\nu) \sum_{j=0}^{\infty}{\frac{(-1)^j}{(2j+1)\,!\,2^{2j+1}}
\frac{\partial^{2j+1} \overline{a^{2}}(X)}{\partial X^{2j+1}}
\frac{\partial^{2j+1} \left[k F(k,X)\right]}{\partial k^{2j+1}}} = 0 \nonumber
\end{eqnarray}
In obtaining (\ref{eq18}) we have expanded $\overline{a^{2}}(x_1)=
\overline{a^{2}}(X+{x\over 2})$ and 
$\overline{a^{2}}(x_2)=\overline{a^{2}}(X-{x\over 2})$ in power series 
around the point $X$, and we have used the relation
$$x^j\,\mathrm{e}^{-\mathbf{i} kx}=(i)^j \frac{\partial^j}{\partial k^j}
\mathrm{e}^{-\mathbf{i} kx}$$
Also we used
$$\lim_{x_2\to x_1}\frac{\partial}{\partial x_1}\rho(x_1,x_2)=\lim_{x\to 0}
\left({1\over 2}\frac{\partial}{\partial X} +\frac{\partial}{\partial x}\right) 
\int_{-\infty}^{+\infty}{\,\mathrm{d}k\, \mathrm{e}^{\mathbf{i} kx} F(k,X)} 
=$$ 
$$={1\over 2}\frac{\partial \overline{a^{2}}(X)}{\partial X}+\mathbf{i} k 
\overline{a^{2}}(X)$$
and similar expressions for the other limits which appear in (\ref{eq14}).

For a linear stability analysis we consider
\begin{equation} \label{eq19}
F(k,X,t)=f(k)+\varepsilon \mathcal{F}(k,X,t)
\end{equation}
and
\begin{equation} \label{eq20}
\overline{a^{2}}(X,t)=\overline{a^{2}_{0}}+\varepsilon \overline{a^{2}_{1}}
(X,t)
\end{equation}
where
\begin{equation} \label{eq21}
\overline{a^{2}_{0}}=\int_{-\infty}^{+\infty}{\,\mathrm{d}k\,f(k)}, 
~~~~~
\overline{a^{2}_{1}}(X,t)=\int_{-\infty}^{+\infty}{\,\mathrm{d}k\,
\mathcal{F}(k,X,t)}
\end{equation}
Here $f(k)$ is the Fourier transform of $\rho_0(|x|)$, which, due to the 
homogenous assumption of the initial state, depends only on $k$, and 
moreover is an even function of $k$.

The linearized kinetic equation in the $(k,X)$ representation is given by
\begin{eqnarray} \label{eq22}
& & \frac{\partial \mathcal{F}}{\partial t}+[k+(\mu+\nu)\overline{a^{2}_{0}}]
\frac{\partial \mathcal{F}}{\partial X}+(\mu+\nu)f\frac{\partial 
\overline{a^{2}_{1}}}{\partial X}+2\nu f\left(\frac{\partial 
\overline{a^{2}_{1}}}{\partial X}\right)^*  \nonumber \\
& & + 4 \lambda \sum_{j=0}^{\infty}\frac{(-1)^j}{(2j+1)\,!\,2^{2j+1}}
\frac{\partial^{2j+1} \overline{a^{2}_{1}}(X)}{\partial X^{2j+1}}
\frac{\partial^{2j+1} f(k)}{\partial k^{2j+1}} \\
& & - 2(\mu+\nu) \sum_{j=0}^{\infty}\frac{(-1)^j}{(2j+1)\,!\,2^{2j+1}}
\frac{\partial^{2j+1} \overline{a^{2}_{1}}(X)}{\partial X^{2j+1}}
\frac{\partial^{2j+1} (kf(k))}{\partial k^{2j+1}} = 0 \nonumber
\end{eqnarray}

The last step is to look for a plane wave solution of (\ref{eq22})
\begin{equation} \label{eq23}
\mathcal{F}(k,X,t)=g_1(k)\mathrm{e}^{\mathbf{i}(QX-\Omega t)}+
g_{2}^{*}(k)\mathrm{e}^{-\mathbf{i}(QX-\Omega^* t)}
\end{equation}
$$\overline{a^{2}_{1}}(X,t)=G_1\mathrm{e}^{\mathbf{i}(QX-\Omega t)}+
G_{2}^{*}\mathrm{e}^{-\mathbf{i}(QX-\Omega^* t)}$$
where
\begin{equation} \label{eq24}
G_i=\int_{-\infty}^{+\infty}{g_i(k)}\,\mathrm{d}k\,,\;\;i=1,2
\end{equation}
Introducing (\ref{eq23}) into (\ref{eq22}) the following equation is found
\begin{eqnarray} \label{eq25}
&& (k-\omega)g_1 + (\mu+\nu)fG_1+2\nu fG_2 \nonumber \\
&& +{{2\lambda}\over Q}\left[f(k+{Q\over 2})-f(k-{Q\over 2})\right]G_1 \\
&& -\frac{\mu+\nu}{Q}\left[h(k+{Q\over 2})-h(k-{Q\over 2})\right]G_1=0 \nonumber
\end{eqnarray}
Here we denoted \ensuremath{\omega = {\Omega\over Q}-(\mu+\nu)
\overline{a^{2}_{0}}} and $h(k)=kf(k)$. Also we used
$$2\sum_{j=0}^{\infty}\frac{1}{(2j+1)\,!}\left({Q\over 2}\right)^{2j+1}
\frac{\partial^{2j+1} f(k)}{\partial k^{2j+1}}=f(k+{Q\over 2})-
f(k-{Q\over 2})$$
and a similar relation for $h(k)$.

In (\ref{eq25}) we have to divide by $(k-\omega)$ and integrate over $k$. 
Introducing the notations
$$I=\int_{-\infty}^{+\infty}{\frac{f(k)}{k-\omega}\,\mathrm{d}k\,}$$
\begin{equation} \label{eq26}
J=\int_{-\infty}^{+\infty}{\frac{f(k+{Q\over 2})-f(k-{Q\over 2})}{k-\omega}\,
\mathrm{d}k\,}
\end{equation}
$$K=\int_{-\infty}^{+\infty}{\frac{h(k+{Q\over 2})-h(k-{Q\over 2})}
{k-\omega}\,\mathrm{d}k\,}$$
the following homogenous algebric linear equation in $G_1$ and $G_2$ is 
obtained
\begin{equation} \label{eq27}
  G_1+(\mu+\nu)IG_1+2\nu IG_2+{{2\lambda}\over Q}JG_1-\frac{\mu+\nu}{Q}KG_1=0
\end{equation}
A second equation is found starting from the complex conjugated of 
(\ref{eq22}), namely
\begin{equation} \label{eq28}
G_2+(\mu+\nu)IG_2+2\nu IG_1+\frac{2\lambda}{Q}JG_2-\frac{\mu+\nu}{Q}KG_2=0
\end{equation}
The compatibility condition for the system (\ref{eq27}),(\ref{eq28}) leads 
us to the following integral stability equations
$$ \label{eqA}
1+(\mu-\nu)I+\frac{2\lambda}{Q}J-\frac{\mu+\nu}{Q}K=0
\eqno\mathrm{(A)}$$
or
$$ \label{eqB}
1+(\mu+3\nu)I+\frac{2\lambda}{Q}J-\frac{\mu+\nu}{Q}K=0
\eqno\mathrm{(B)}$$
In the next section these will be solved for different initial conditions 
$f(k)$.

\section{Solution of the integral stability equation}

The first distribution function we shall discuss is a $\delta$-spectrum
\begin{equation} \label{eq29}
f(k)=\overline{a^{2}_{0}}\delta(k)
\end{equation}
It doesn't describe a realistic situation as it coresponds to a constant 
two-point correlation function in the initial state, 
$\rho_0=\overline{a^{2}_{0}}=const.$. But the calculations are easily 
done and can be considered as a limit case to which more realistic situations 
can be compared. The integrals $I,J,K$ are immediatly calculated
\begin{equation} \label{eq30}
I=-\frac{\overline{a^{2}_{0}}}{\omega},\;J=\frac{\overline{a^{2}_{0}}Q}
{\omega^2 - {{Q^2}\over 4}},\;K=0
\end{equation}
Then both integral equations (A) and (B) become an algebraic equation of 
third order in $\omega$
\begin{equation} \label{eq31}
\omega^3 -\bar{\mu} \overline{a^{2}_{0}}\omega^2+
(2\lambda\overline{a^{2}_{0}}-{{Q^2}\over 4})\omega +
\bar{\mu}\overline{a^{2}_{0}}{{Q^2}\over 4}=0
\end{equation}
where $\bar{\mu}=\mu_A=\mu-\nu$ for A-equation and 
$\bar{\mu}=\mu_B=\mu+3\nu$ for B-equation. With 
$\omega=y+{1\over 3}\bar{\mu}\overline{a^{2}_{0}}$ 
it is reduced to the canonical form
\begin{equation} \label{eq32}
y^3+py+q=0
\end{equation}
where
$$p=2\lambda\overline{a^{2}_{0}}-{1\over 3}(\bar{\mu}\overline{a^{2}_{0}})^2
-{{Q^2}\over 4}$$
\begin{equation} \label{eq33}
q={2\over 3}\bar{\mu}\overline{a^{2}_{0}}\left[\lambda
\overline{a^{2}_{0}}-{1\over 9}(\bar{\mu}\overline{a^{2}_{0}})^2+
{{Q^2}\over 4}\right]
\end{equation}
Complex solutions of (\ref{eq32}) are obtained if the discriminant 
$\Delta=\left({p\over 3}\right)^3+\left({q\over 2}\right)^2$ 
is positive. The sign of $\Delta$ is controlled by the sign of $p$. For 
$\lambda\neq 0$ we have $p>0$ if 
$Q^2/4<2\lambda\overline{a^{2}_{0}}-{1\over 3}(\bar{\mu}
\overline{a^{2}_{0}})^2$, corresponding to the long wavelength region. 
With the notations 
\begin{equation} \label{eq35}
\sqrt{\frac{p^3}{27}}={q\over 2}\tan\alpha,\;\;\tan\varphi=
\sqrt[3]{\tan{\alpha\over 2}}
\end{equation}
we have
\begin{equation} \label{eq36}
\mathrm{Im}\,\omega = \frac{\sqrt{p}}{\sin2\varphi}
\end{equation}
But if $p<0$ we can still have $\Delta>0$. Writing
\begin{equation} \label{eq37}
{{Q^2}\over 4}=2\lambda\overline{a^{2}_{0}}-{1\over 3}(\bar{\mu}
\overline{a^{2}_{0}})^2+3\tilde{Q}^2
\end{equation}
the condition $\Delta>0$ leads us to the inequality
\begin{equation} \label{eq38}
\tilde{Q}^3-\bar{\mu}\overline{a^{2}_{0}}\tilde{Q}^2<\bar{\mu}
\overline{a^{2}_{0}}\left[\lambda\overline{a^{2}_{0}}-{4\over 27}
(\bar{\mu}\overline{a^{2}_{0}})^2\right]
\end{equation}
In this case ($\Delta>0$ and $p<0$) the usual notations are
$$ \label{eq35p}
\sqrt{-\frac{p^3}{27}}={q\over 2}\sin \alpha,\;\;\tan\varphi=
\sqrt[3]{\tan{\alpha\over 2}}
\eqno(\ref{eq35}')$$
and one obtaines
$$ \label{eq36p}
\mathrm{Im}\,\omega=\sqrt{-p}\cot 2\varphi
\eqno(\ref{eq36}')$$
The instability remains in the long wavelength region.

When $\lambda=0$ the situation is completely changed. Then
$$p=-\left[{{Q^2}\over 4}+{1\over 3}(\bar{\mu}\overline{a^{2}_{0}})^2\right]$$
\begin{equation} \label{eq39}
q={2\over 3}\mu\overline{a^{2}_{0}}\left[{{Q^2}\over 4}-{1\over 9}
(\bar{\mu}\overline{a^{2}_{0}})^2\right]
\end{equation}
and $\Delta>0$ can be realized only if $Q>{2\over 3}\bar{\mu}
\overline{a^{2}_{0}}$. This means that the instability develops for 
higher wave vectors $Q$, a result in complete agreement to the conclusions 
of the DAMI analysis of the derivative NLS equations. Using the 
same notations ($34'$) we get the same result ($35'$), and the requirement 
$\Delta>0$ reduces to
\begin{equation} \label{eq040}
\left({q\over 2}\right)^2 >-\left({p\over 3}\right)^3
\end{equation}
an inequality from which the limit value of ${{Q^2}\over 4}$ can be calculated.

A more realistic situation is a Lorentzian distribution
\begin{equation} \label{eq40}
f(k)={\overline{a^{2}_{0}}\over \pi}\frac{\sigma}{k^2 + \sigma^2}
\end{equation}
which corresponds to an exponential decaying two-point correlation 
function in the initial state
\begin{equation} \label{eq41}
\rho_0(x)=\overline{a^{2}_{0}} \mathrm{e}^{-\sigma x}\; x>0
\end{equation}
The integrals {\bf I}, {\bf J}, {\bf K} are easily calculated in the 
$k$-complex plane assuming $\omega$ complex with $\mathrm{Im}\,\omega >0$. 
We obtain
$$
I=-\frac{\overline{a^{2}_{0}}}{\tilde \omega}
$$
\begin{equation} \label{eq42}
J=\frac{\overline{a^{2}_{0}} Q}{\tilde \omega^2 -{Q^2 \over 4}}
\end{equation}
$$
K=\frac{-\mathbf{i} \sigma \overline{a^{2}_{0}} Q}{\tilde \omega^2 -
{Q^2 \over 4}}
$$
where $\tilde \omega=\omega+\mathbf{i} \sigma$, and the instability 
condition $\mathrm{Im}\,\omega >0$ becomes for the new variable 
$\tilde \omega$, $\mathrm{Im}\,\tilde\omega>\sigma$.

To have a point of reference, let us recall the result for the usual 
NLS equation ($\mu=\nu=0$) (Visinescu et al 2003, Grecu D. et 
al (2004, 2005), Grecu A.T. 2005)
\begin{equation} \label{eq43}
\mathrm{Im}\,\omega=\sqrt{2\lambda \overline{a^{2}_{0}} -{Q^2 \over 4}}-
\sigma
\end{equation}
We get $\mathrm{Im}\,\omega>0$ if
\begin{equation} \label{eq44}
Q<2\sqrt{2\lambda\overline{a^{2}_{0}}-\sigma^2}
\end{equation}
which shows that the MI is restricted to the long wave-length region and 
is strongly dependent on the correlation length in the initial state; 
if $\sigma^2>2\lambda\overline{a^{2}_{0}}$, the initial state is stable 
to small modulations.

When (\ref{eq42}) are introduced in (A) and (B), we get again an algebraic 
equation of third order in $\tilde \omega$, but now with complex 
coefficients. It can be reduced to the canonical form with
\begin{equation} \label{eq45}
p=2\left(\lambda+{\mathbf{i}\over 2}\overline{\nu}\sigma\right)
\overline{a^{2}_{0}}-{1\over 3}(\overline{\mu}\overline{a^{2}_{0}})^2-
{Q^2\over 4}
\end{equation}
$$
q={2\over 3}\overline{\mu}\overline{a^{2}_{0}}\left[\left(\lambda+
{\mathbf{i}\over 2}\overline{\nu}\sigma\right)\overline{a^{2}_{0}}-
{1\over 9}(\overline{\mu}\overline{a^{2}_{0}})^2+{Q^2\over 4}\right]
$$
where $\overline{\mu}$ has the same meaning as in the case of the 
$\delta$-spectrum, and $\overline{\nu}=\mu+\nu$. As mentioned before the 
MI region coresponds to $\mathrm{Im}\,\tilde\omega > \sigma$.

\begin{figure}[ht]
\label{fig:3}
\begin{center}
\includegraphics[width=5in]{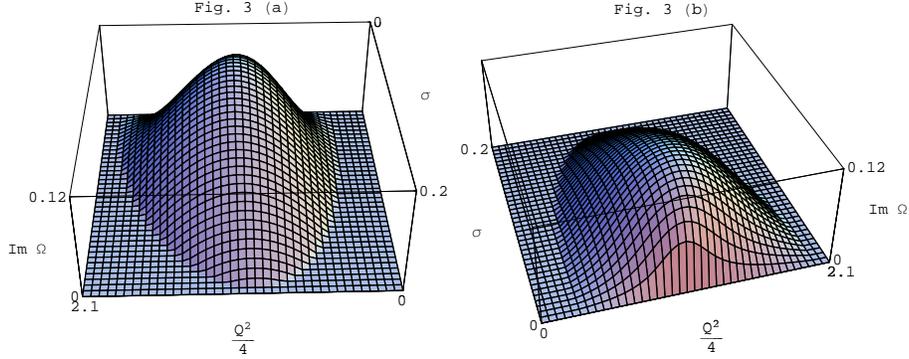}
\end{center}
\caption{\small Instability region for dNLS I. Three dimensional 
plot of $\mathrm{Im}\,\Omega$ versus 
$Q^2/4$ and $\sigma$; Two different orientantions.}
\end{figure}

Instead of working on the general case let us consider only the situations 
of the dNLS equations. We shall start with the dNLS-II equation 
($\lambda=\nu=0$). Then both equations (A) and (B) reduce to a single 
equation
\begin{equation} \label{eq46}
1-\mu\overline{a^{2}_{0}}{1\over \tilde{\omega}}+\mathbf{i}\mu
\overline{a^{2}_{0}}\frac{\sigma}{\tilde{\omega^2}-{Q^2\over 4}}=0
\end{equation}
It is convenient to measure $\tilde\omega,\,\sigma\,\mathrm{and}\,Q$ 
in units of $|\mu|\overline{a^{2}_{0}}$. If $\varepsilon=\mathrm{sgn}\,\mu$, 
the following algebraic equation of third order is found:
\begin{equation} \label{eq47}
\tilde \omega^3-\varepsilon\tilde \omega^2 -\left({Q^2\over 4}-
\mathbf{i}\varepsilon\sigma\right)\tilde \omega+\varepsilon{Q^2\over 4}=0
\end{equation}
It has a certain symmetry having solutions of the form $\tilde \omega=
\varepsilon\tilde \omega_r+\mathbf{i}\tilde \omega_i$. As we are interested 
in the imaginary part of these solutions, the result is independent on 
the sign of $\mu$. The equation (\ref{eq47}) is solved numerically. 
Measuring $\Omega$ in units of $(\mu\overline{a^{2}_{0}})^2$, one has 
$\mathrm{Im}\,\Omega=Q(\mathrm{Im}\,\tilde \omega -\sigma)$. A three 
dimensional plot , $\mathrm{Im}\,\Omega$ function of $\sigma$ and $Q^2/4$ 
is given in Fig. \ref{fig:3}  As one sees, the neighbourhood of $Q=0$ 
is excluded from the instability region, which extends on a finite domain 
of $Q$ and $\sigma$. One can find a limit value for 
$\sigma\,(\sigma_C\simeq0.194)$ and the MI exists only for $\sigma<\sigma_C$.

Now let us consider the case of dNLS-I equation ($\lambda=0,\,\mu=\nu$). 
The equation (A) becomes
\begin{equation} \label{eq48}
1+2\mathbf{i}\frac{\mu\overline{a^{2}_{0}}}{\tilde \omega^2-{Q^2\over 4}}=0
\end{equation}
which is easily solved giving
\begin{equation} \label{eq49}
\mathrm{Im}\,\tilde \omega={1\over\sqrt{2}}\left[\sqrt{\left({Q^2\over 4}
\right)^2+4(\mu\overline{a^{2}_{0}})^2}-{Q^2\over 4}\right]^{1/2}
\end{equation}
and the MI region is determined by
\begin{equation} \label{eq50}
Q^2<4\left[\left(\frac{\mu\overline{a^{2}_{0}}}{\sigma}\right)^2-
\sigma^2\right],\;\sigma^2<\mu\overline{a^{2}_{0}}
\end{equation}
This instability region is located near the origin in the plane 
$(Q,\,\sigma)$.

Another instability region results from (B), which writes
\begin{equation} \label{eq51}
\tilde \omega^3-4\varepsilon\tilde \omega^2-\left({Q^2\over 4}-
2\mathbf{i}\varepsilon\sigma\right)\tilde \omega+4\varepsilon{Q^2\over 4}=0
\end{equation}
where as in the case of dNLS-II, we measure $\tilde \omega,\,\sigma$ 
and $Q$ in units of $|\mu|\overline{a^{2}_{0}}$. 
This equation has the same form as (\ref{eq46}). The instability region 
is obtained numerically and is represented in Fig. 4. Again the 
$Q\simeq 0$ region is excluded and one finds a critical value 
$\sigma_c\simeq 0.441$ for $\sigma$, which determines the superior limit 
of the MI region.
\begin{figure}[ht]
\label{fig:4}
\begin{center}
\includegraphics[width=5in]{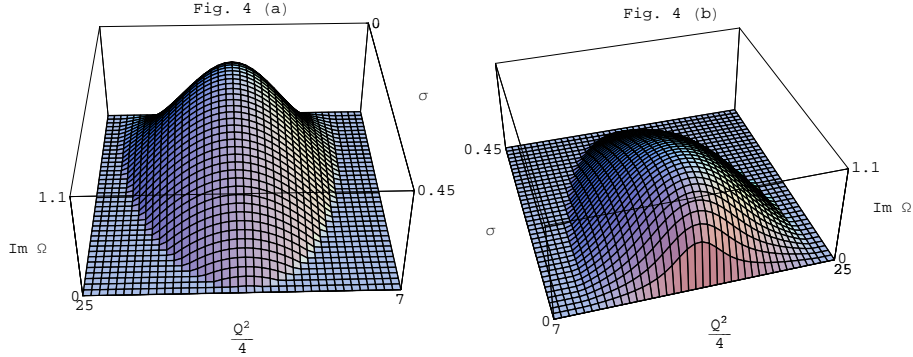}
\end{center}
\caption{\small Instability region for dNLS II. Three dimensional plot 
of $\mathrm{Im}\,\Omega$ versus $Q^2/4$ and $\sigma$; Two different 
orientantions.}
\end{figure}

\section{Conclusions}

Several conclusions result from this analysis. For the extended NLS 
equation (\ref{eq4}) the instability region in a DAMI analysis is 
dependent on the wave-vector $\vec{k}$ of the carrying wave. In the 
case of derivative NLS equations the long wave-length region is either 
forbidden, or highly unfavourable for the development of MI.

In the SAMI instead of the $k$-dependence the instability region is 
now dependent on the correlation length in the initial 2-point correlation 
function. Although we got explicit results only for a $\delta-$ and 
Lorentzian distribution function we conclude that the MI is possible 
only for long-range correlations in the initial state. In the case of a 
$\delta$-function initial distribution, in a limit situation corresponding 
to an infinite correlation length, the results are similar with those 
obtained from DAMI analysis; for the dNLS equations again the long 
wave-length region is highly unfavourable. The same conclusions can be 
drawn for the more realistic case of a Lorentzian initial distribution 
function, namely that the MI region is restricted to a finite region of 
($Q,\;\sigma$) plane containing the axis $\sigma = 0$. A critical value 
$\sigma_c$ is determined for both dNLS equations and again the $Q \simeq 0$ 
region is unfavourable for the instability development.\\
Another realistic initial condition is a Gaussian distribution function
$$
f(k)=\overline{a^{2}}{0} {1 \over\sqrt{2\pi}\sigma}\mathrm{e}^
{-\frac{k^2}{2\sigma^2}}
$$
Calculations done only for NLS eq. (see Alber 1978) reveal the same 
conclusions, namely the MI is possible only for long-range 
correlations in the initial state ($\sigma$ small) and in the 
long wave-length region. The result is more complicated, being 
expressed through some special functions (complex integral function 
$\omega(z)$, Abramowitz and Stegun 1965). 

Such calculations for dNLS equations are not yet performed.

The influence of the statistical properties of the medium on the 
development of large amplitude waves, is a problem of special interest 
in hydrodynamics, especially for the generation of freak waves in 
the ocean. In a linear theory there is no interaction between ocean 
waves. A focusing phenomenon of wave energy may occur only when 
constructing interference takes place. The situation is competely 
different when nonlinear wave-wave interaction is present. Then a 
wave can borrow energy from its neighbours and due to this extra-focusing 
phenomenon, waves with very high amplitude can be generated. In this 
process the knowledge of the initial conditions of the oceanic waves 
is essential. Accordingly, many experimental works have been conducted. 
For a long time the Joint North Sea Wave Project (JONSWAP) has studied 
the power spectrum of the oceanic waves (Komen et al 1994), resulting a 
non-Gaussian expression. It was used to calculate the probability to 
generate large amplitude oceanic waves (analytically and numerically) 
in given realistic conditions (Janssen 1983, Onorato et al 2001, 
2003, 2004).

Calculations using other initial distribution functions, reflecting 
different physical situations, would be of real interest 
for sure, even in the case of derivative NLS equations.

\noindent\underline{\bf Acknowledgements}\\

The present work was done under the contract CEX-D11-9 with the Ministry 
of Education and Research from Romania.

\section{References}

Abdulaev, F.Kh., Darmanyan, S.A., Garnier, J. (2002) in 
{\it Progress in Optics} {\bf 44}, 303 (ed. E. Wolf, Elsevier)
\\
Abramowitz, M., Stegun, I., (1965) {\it Handbook of Mathematical 
Functions} (National Bureau of Stand)
\\
Alber, I.E. (1978) {\it Proc. Roy. Soc. London A} {\bf 363}, 525 
\\
Benjamin, T.B., Feir, J.E., (1967) {\it J. Fluid Mech.} {\bf 27}, 
417
\\
Bespalov, V. I., Talanov, V.I., (1966) {\it Prisma JETP} {\bf 3}, 
417
\\ 
Chen, H.H., Lee, Y.C., (1979) {\it Physica Scripta} {\bf 20}, 490 
\\
Dodd, R.K., Eilbeck, J.C., Gibbon, J.D., Morris, H.C., (1982) {\it 
Solitons and Nonlinear Wave Equations} (Acad. Press, New York)
\\
Fedele, R., Anderson, D., (2000) {\it J. Opt. B: Quantum Semiclassical 
Optics} {\bf 2}, 207 
\\
Fedele, R., Anderson, D., Lisak, M., (2000) {\it Physica Scripta 
T} {\bf 84}, 27 
\\
Fedele, R., Shukla, P.K., Onorato, M., Anderson, D.,  
Lisak, M., (2002) {\it Phys. Rev. Lett. A} {\bf 303}, 61 
\\
Grecu, A.T., Grecu, D., Visinescu, A., (2005) {\it Rom. J. 
Phys.} {\bf 50}, 127 
\\
Grecu, A.T., (2005) {\it Ann. Univ. Craiova, Physics AUC} {\bf 15}
(part I), 177 
\\
Grecu, D., Visinescu, A., (2004) in {\it Nonlinear Waves. Classical 
and Quantum Aspects}, p. 151 (eds. F. Kh. Abdulaev, V. V. Konotop, 
Kluwer Acad. Publ.)
\\
Grecu, D., Visinescu, A., (2005) {\it Theor. Math. Phys.} {\bf 144}, 
927 
\\
Grecu, D., Visinescu, A., (2005)  {\it Rom. J. Phys.} {\bf 50}, 137 
\\
Hall, B., Lisak, M., Anderson, D., Fedele, R., Semenov, V.E., 
(2002) {\it Phys. Rev. E} {\bf 65}, 035602 R 
\\
Janssen, P.A.E.M., (1983) {\it J. Fluid Mech.} {\bf 133}, 113 
\\
Kaup, D.J., Newell, A.C., (1978) {\it J. Math. Phys.} {\bf 19}, 798 
\\
Komen, J.C., Cavaleri, L., Donelan, M., Hasselman, K.,  
Hasselman, S.,  Jansen, P.A.E.M., (1994) {\it Dynamics and Modelling of 
Ocean Waves} (Cambridge Univ. Press) 
\\
Lisak, M., Hall, B., Anderson, D.. Fedele, R., Semenov, V.E., Shukla, 
P.K., Hasegawa, A., (2002) {\it Physics Scripta T} {\bf 98}, 12 
\\
Mio, K., Ogino, T., Minami, K., Takeda, S., (1976) {\it J. Phys. Soc. 
Japan} {\bf 41}, 265, 667 
\\
Mj\o lhus, E., (1976) {\it J. Plasma Physics} {\bf 16}, 321 
\\
Moyal, J.E., (1949) {\it Proc. Cambridge Phyl. Soc.} {\bf 45}, 99 
\\
Nakamura, A., Chen, H.H., (1980) {\it J. Phys. Soc. Japan} {\bf 49}, 813 
\\
Onorato, M., Osborne, A.R., Serio, M., Bertone, S., (2001) 
{\it Phys. Rev. Lett.} {\bf 86}, 5831 
\\
Onorato, M., Osborne, A., Fedele, R., Serio, M., (2003) {\it Phys. 
Rev. E} {\bf 67}, 046305 
\\
Onorato, M., Osborne, A.R., Serio, M., Cavaleri, L., Brandini, 
C., Stansberg C.T., (2004) {\it Phys. Rev. E} {\bf 70}, 067302 
\\
Remoissenet, M., (1999) {\it Waves Called Solitons} (Springer Verlag, 
Berlin) 
\\
Scott, A., (2003) {\it Nonlinear Science. Emergence and Dynamics of 
Coherent Structures} (Oxford Univ. Press) 
\\
Visinescu, A., Grecu, D., (2003) {\it Eur. Phys. J. B} {\bf 34}, 225 
\\
Visinescu, A., Grecu, D., (2003) {\it Rom. J. Phys.} {\bf 48}, 787 
\\
Toda, M., Kubo, R., Saito, N., (1995) {\it Statistical 
Physics} vol. 1 (Springer, Berlin) 
\\
Wigner, E., (1932) {\it Phys. Rev.} {\bf 40}, 749 

\end{document}